\documentclass[useAMS,usenatbib,usegraphicx]{mn2e}
\usepackage{times}
\usepackage{booktabs}

\usepackage{color}

\title[Dust evolution in high-$z$ quasars]
{Evolution of grain size distribution in high-redshift dusty quasars:
Integrating large amounts of dust and unusual extinction curves}
\author[Nozawa et al.]{Takaya Nozawa,$^{1}$\thanks{E-mail:
takaya.nozawa@nao.ac.jp}
Ryosuke S. Asano,$^{2}$
Hiroyuki Hirashita$^{3}$
and Tsutomu T. Takeuchi$^{2}$
\\
$^{1}$National Astronomical Observatory of Japan, Mitaka, Tokyo
181-8588, Japan \\
$^{2}$Department of Particle and Astrophysical Science, Nagoya 
University, Furo-cho, Chikusa-ku, Nagoya 464-8602, Japan\\
$^{3}$Institute of Astronomy and Astrophysics, Academia Sinica, P.\ O.\
Box 23-141, Taipei 10617, Taiwan}

\begin{document}

\date{\today}

\pagerange{\pageref{firstpage}--\pageref{lastpage}} \pubyear{2014}

\maketitle

\label{firstpage}

\begin{abstract}
The discoveries of huge amounts of dust and unusual extinction curves 
in high-redshift quasars ($z \ga 4$) cast challenging issues on the 
origin and properties of dust in the early universe.
In this Letter, we investigate the evolutions of dust content and 
extinction curve in a high-$z$ quasar, based on the dust evolution 
model taking account of grain size distribution.
First, we show that the Milky-Way extinction curve is reproduced by 
introducing a moderate fraction ($\simeq$0.2) of dense molecular-cloud 
phases in the interstellar medium for a graphite-silicate dust model.
Then we show that the peculier extinction curves in high-$z$ quasars 
can be explained by taking a much higher molecular-cloud fraction 
($\ga$0.5), which leads to more efficient grain growth and coagulation, 
and by assuming amorphous carbon instead of graphite.
The large dust content in high-$z$ quasar hosts is also found to be a 
natural consequence of the enhanced dust growth.
These results indicate that grain growth and coagulation in molecular 
clouds are key processes 
that can increase the dust mass and change the size distribution of 
dust in galaxies, and that, along with a different dust composition, 
can contribute to shape the extinction curve.

\end{abstract}


\section{Introduction}

There are some pieces of evidence that the extinction curves observed 
for high-redshift ($z \ga 4$) quasars are different from those for the 
lower-redshift counterparts.
Maiolino et al.\ (2004) found that the quasar SDSS J1048+4637
(hereafter J1048+4637) at $z=6.2$ exhibits an unusual UV extinction 
curve, being flat at wavelengths $\lambda \ga 1700$ \AA~and rising at 
$\lambda \la 1700$ \AA.
Gallerani et al.\ (2010) showed that the UV extinction curves in 
seven reddened quasars at $z =$ 3.9--6.4 have no 2175 \AA~bump and are
flatter than those in quasars at $z \la 2$ which are described by the 
extinction curve in the Small Magellanic Cloud (SMC) 
(Richards et al.\ 2003; Hopkins et al.\ 2004).
It has also been suggested that the extinction laws in the quasars 
CFHQS J1509--1749 at $z=6.12$ (Willott et al.\ 2007) and SDSS 
J1044--0125 at $z=5.8 $ (Maiolino et al.\ 2004) are similar to that 
in the quasar J1048+4637 (but see Hjorth et al.\ 2013).
The UV extinction curve in J1048+4637 is successfully reproduced by 
the models of dust produced by Type II supernovae (SNe II) 
(Maiolino et al.\ 2004; Hirashita et al.\ 2005;
Bianchi \& Schneider 2007; Hirashita et al.\ 2008), which seems to 
support the idea that the interstellar dust in such early epochs was 
predominantly supplied by SNe II generating from massive stars with 
short lifetimes.

On the other hand, a significant fraction of high-$z$ quasars,
including J1048+4637, manifest the presence of dust mass in excess of 
$10^8$ $M_\odot$ in their host galaxies (e.g.\ Calura et al.\ 2014).
Such huge amounts of dust grains can be explained by dust production 
in SNe II if one SN ejects more than 1 $M_\odot$ of dust
(Dwek, Galliano, \& Jones 2007; Dwek \& Cherchneff 2011) and/or 
if the stellar initial mass function (IMF) is biased to a much higher 
mass than the local one 
(Valiante et al.\ 2009; Gall, Andersen, \& Hjorth\ 2011a, 2011b).
However, many studies have argued that, in addition to the 
contributions from SNe II and asymptotic giant branch (AGB) stars, 
grain growth via accretion of gaseous metals in the interstellar 
medium (ISM) is required to account for the observed dust mass
(Michalowski et al.\ 2010; Pipino et al.\ 2011; Mattsson 2011; 
Valiante et al.\ 2011; Kuo \& Hirashita 2012).
Thus, if grain growth is the dominant process for increasing the dust 
mass, as is considered in the Milky Way (MW) 
(Dwek \& Scalo 1980; Zhukovska, Gail, \& Trieloff 2008; Inoue 2011; 
Hirashita \& Kuo 2011; Asano et al.\ 2013a), then the observed unusual 
extinction curves may not necessarily reflect the SN origin of dust in 
high-$z$ quasars.

Most of the dust evolution models focused only on reproducing the 
observed dust mass by estimating the contributions of dust formation 
and destruction to the total dust budget in galaxies.
These models assumed a representative grain size or a specific grain 
size distribution throughout the galaxy evolution, and thus cannot 
make any prediction on extinction curves.
In addition, the efficiencies of dust destruction and grain growth,
as well as the efficiency of extinction by dust, heavily depend on
the size distribution of the dust grains.
In particular, the efficiency of grain growth is higher for a smaller 
surface-to-volume ratio of the grains, and the efficient production 
of small grains through shattering by grain-grain collisions is found 
to be necessary for boosting the mass of dust via grain growth 
(Hirashita \& Kuo 2011; Kuo \& Hirashita 2012; Asano et al.\ 2013b).
Hence, to model properly the evolution of dust mass and to predict 
the extinction curves, it is essential to clarify how the size 
distribution of grains changes in the course of galaxy evolution.

Recently, Asano et al.\ (2013b) constructed an evolution model of 
grain size distribution taking into account the fundamental physical 
processes of formation and destruction of dust.
This model enables us to predict the extinction curves in galaxies
as a function of time (Asano et al.\ 2014).
In this Letter, we apply their dust evolution model to investigate 
the evolution of grain size distribution and the expected extinction 
curves in high-$z$ quasars.\footnote{
As in the previous studies, we here assume that the extinction curves 
of high-$z$ quasars are shaped by the interstellar dust 
in the host galaxies.
However, it may be possible that they reflect the 
properties of dust in the local environments such as the nuclear 
tori (e.g., Gaskwell et al.\ 2004).}
In Section 2, we briefly describe our dust evolution model, and 
illustrate how the MW extinction curve is reproduced by our model.
In Section 3, we present the results for the evolution of dust and 
extinction curves in high-$z$ dusty quasars.
In Section 4, we provide the conclusions.
Throughout this Letter, we assume dust grains to be spherical.

\section{Dust evolution model and reproduction of extinction curve 
in the Milky Way} 
\label{sec:model}

\subsection{Review of the adopted model}

In this study, we adopt the evolution model of dust in galaxies 
constructed by Asano et al.\ (2013b).
This model self-consistently treats the time variation of size 
distribution of interstellar dust caused by the following dust 
processes:
dust formation by SNe II and AGB stars, 
grain growth due to the accretion of gaseous metals, 
dust destruction by interstellar shocks, 
shattering and coagulation due to grain-grain collisions in the ISM.

In the model, the composition, size distribution, and mass of dust 
ejected by SNe II are taken from the calculations of dust formation 
in the ejecta and dust destruction in the reverse shocks by 
Nozawa et al.\ (2003, 2007).
The masses of various grain species produced by AGB stars are taken 
from Zhukovska et al.\ (2008), assuming that the size distributions 
of those grains are lognormal with a peak at a grain radius 
$a= 0.1$ $\mu$m.
The initial stellar mass ranges of SNe II and AGB stars are set to 
be 8--40 $M_\odot$ and 1--7 $M_\odot$, respectively, under the Salpeter 
IMF in the stellar mass range of 0.1--100 $M_\odot$.
The star formation rate at a time $t$ is calculated as  
$\mbox{SFR}(t) = M_{\rm ISM}(t)/\tau_{\rm SF}$, where 
$M_{\rm ISM}(t)$ is the total mass of gas and dust in the ISM, and 
$\tau_{\rm SF}$ is the timescale of star formation.

In this dust evolution model, the size distributions of grain species 
except for carbonaceous grains are summed up so that they are 
integrated as silicate grains.
This corresponds to the fact that two dust species, carbonaceous 
(graphite) and silicate grains, are taken as major components of 
interstellar dust, to simplify the calculations of grain growth, 
shattering, and coagulation, as well as the calculations of 
extinction curves.
For the growth process of grains, the accretion of refractory elements 
onto pre-existing grains is calculated with 
the sticking probability of unity, by following the prescription in 
Hirashita \& Kuo (2011).
As for the destruction of interstellar dust in high-velocity shocks 
driven by supernovae, the modification of grain size distribution by 
sputtering is computed on the basis of the formulae in 
Yamasawa et al.\ (2011).

The model also takes into account the processing of dust through 
shattering and coagulation in the interstellar turbulence 
(Hirashita \& Yan 2009).
It is assumed that shattering occurs if the relative velocity of grains 
is higher than 1.2 and 2.7 km s$^{-1}$ for carbonaceous and silicate 
grains, respectively (Jones, Tielens, \& Hollenbach 1996).
Coagulation occurs for the relative velocities lower than 
$10^{-3}$--$10^{-1}$ km s$^{-1}$, depending on the composition and 
radius of grains (Chokshi, Tielens, \& Hollenbach 1993).
Grain velocities in turbulence are taken from the results of 
magnetohydrodynamic calculations by Yan, Lazarian, \& Draine (2004).

In order to follow the contributions of shattering and coagulation 
in different phases of the ISM, multiple phases are considered within 
a framework of the one-zone closed model;
the contributions of dust processes in each phase are weighted
by a mass fraction of each ISM phase $\eta_i$, where $i$ denotes 
the phase of the ISM.
Asano et al.\ (2013b) considered a two-phase model of the ISM: 
warm neutral medium (WNM, with gas temperature $T_{\rm gas}=6000$ K,
and hydrogen number density $n_{\rm H}=0.3\;{\rm cm}^{-3}$) and 
cold neutral medium (CNM, $T_{\rm gas}=100$ K, $n_{\rm H}=30\;{\rm cm}^{-3}$)
with $\eta_{\rm WNM} = \eta_{\rm CNM} = 0.5$.
In this two-phase model, grain growth, whose rate is proportional to 
the gas density, occurs only in the CNM.
Shattering due to collisions between grains occurs in the WNM, 
whereas coagulation due to collisions between grains occurs 
predominantly in the CNM.

We also calculate the extinction curves following Asano et al.\ (2014), 
who used the optical constants of graphite and silicate taken from 
Draine \& Lee (1984).
In this Letter, we also adopt the optical constants of amorphous 
carbon (see Section 3).

\subsection{New features in this work}

Asano et al.\ (2014) found that the extinction curve derived from 
the dust evolution model by Asano et al.\ (2013b) shows too steep a
UV slope and too large a 2175 \AA~bump to be consistent with that 
in the MW (see Figure 6 in Asano et al.\ (2014)).
This indicates that the grain size distribution predicted by this 
model is largely overabundant in small grains, compared with the 
interstellar dust models deduced from the fitting to the average 
MW extinction curve (Mathis, Rumpl, \& Nordsieck 1977; 
Weingartner \& Draine 2001; Zubko, Dwek, \& Arendt 2004).
The classical interstellar dust model by Mathis et al.\ (1977) has 
a power-law size distribution in the range of $a =$ 0.005--0.25 
$\mu$m, while the grain size distribution by Asano et al.\ (2013b) 
has a bump at $a =$ 0.01--0.03 $\mu$m.

Asano et al.\ (2014) argued that one of the reasons for this 
overabundance of small grains (underabundance of large grains) might
be inefficient coagulation resulting from the unreasonably low 
threshold velocities of coagulation. 
Thus, they also examined a case where the threshold velocity of 
coagulation is removed, keeping the other parameters unchanged 
($\tau_{\rm SF} = 5$ Gyr and $\eta_{\rm WNM} = \eta_{\rm CNM} = 0.5$).
However, as shown in Figure 1, the resulting extinction curve at 
$t=10$ Gyr is still too steep with too strong a UV bump, although 
it is nearer to the observed MW extinction curve than the original 
model with the coagulation threshold.
The absence of coagulation threshold is supported by the studies of 
the variation of extinction curves in the MW 
(Hirashita \& Voshchinnikov 2014).

\begin{figure}
\centering\includegraphics[width=0.42\textwidth]{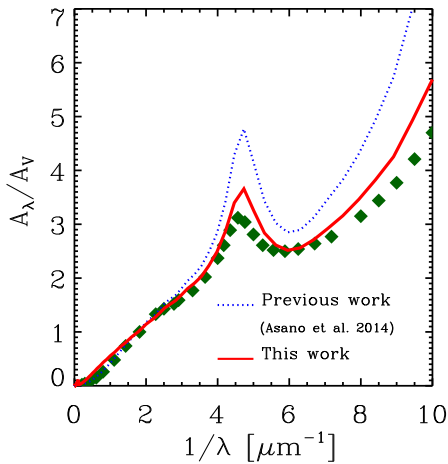}
\caption{
Extinction curves at a galactic age of $t=10$ Gyr obtained from the 
calculations of the dust evolution with no threshold velocity of 
coagulation for the star formation timescale of $\tau_{\rm SF}=5$ Gyr.
The dotted line is the result for 
$\eta_{\rm WNM} = \eta_{\rm CNM} = 0.5$, which was discussed in 
Asano et al.\ (2014). 
The solid line shows the extinction curve for
$\eta_{\rm WNM} = 0.5$, $\eta_{\rm CNM} = 0.3$, and $\eta_{\rm MC} = 0.2$,
which is examined in this study.
The filled diamonds denote the average extinction curve in the MW 
(Whittet 2003).
}
\label{fig1}
\end{figure}

To overcome this defect, in the present calculations, we include the 
third phase of the ISM, molecular clouds (MC, $T_\mathrm{gas}=25$ K, 
$n_\mathrm{H}=300$ cm$^{-3}$), where the gas density is high enough to 
increase the efficiency of coagulation.
In fact, a three-phase ISM model is suggested, for which the MC phase 
has a similar mass fraction to the CNM (e.g.\ Whittet 2003).
The solid line in Figure 1 shows the extinction curve for the
three-phase ISM with $\eta_{\rm WNM} = 0.5$, $\eta_{\rm CNM} = 0.3$, 
and $\eta_{\rm MC} = 0.2$.
The extinction curve becomes flatter thanks to the enhanced 
production of large grains by coagulation (and grain growth) in MCs, 
and much better agrees with the MW extinction curve, although the 
agreement is not perfect.
Nevertheless, given the variety of extinction curves along the lines
of sight (Fitzpatrick \& Massa 2007; Nozawa \& Fukugita 2013), 
we regard that the agreement is satisfactory.
This allows us to conclude that our dust evolution model can reproduce 
the MW extinction curve with a reasonable choice of $\eta_i$ if the 
coagulation threshold velocity is excluded.
In what follows, we assume no threshold velocity for the calculation 
of coagulation.

\begin{figure*}
\centering\includegraphics[width=0.42\textwidth]{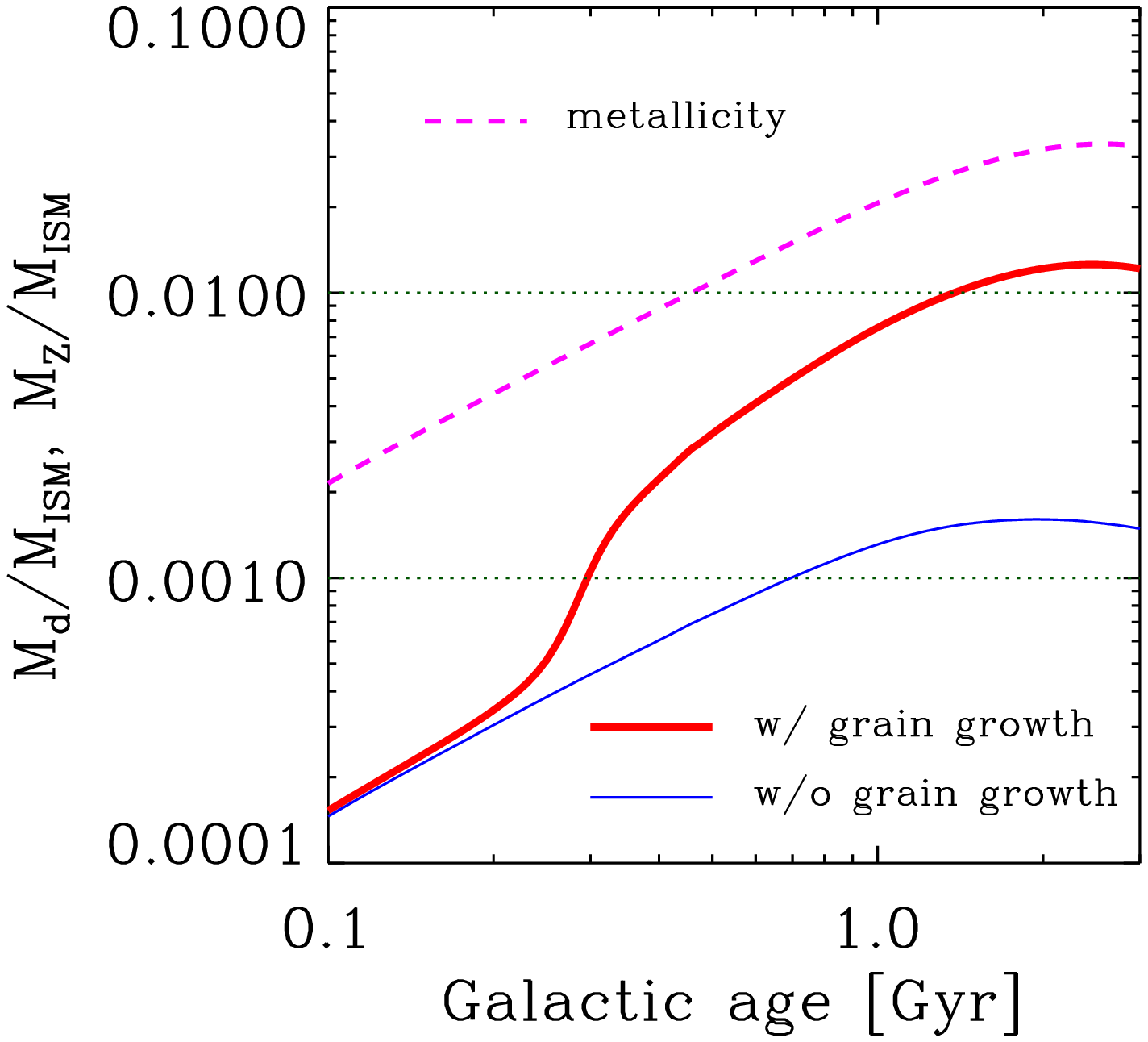}
\centering\includegraphics[width=0.42\textwidth]{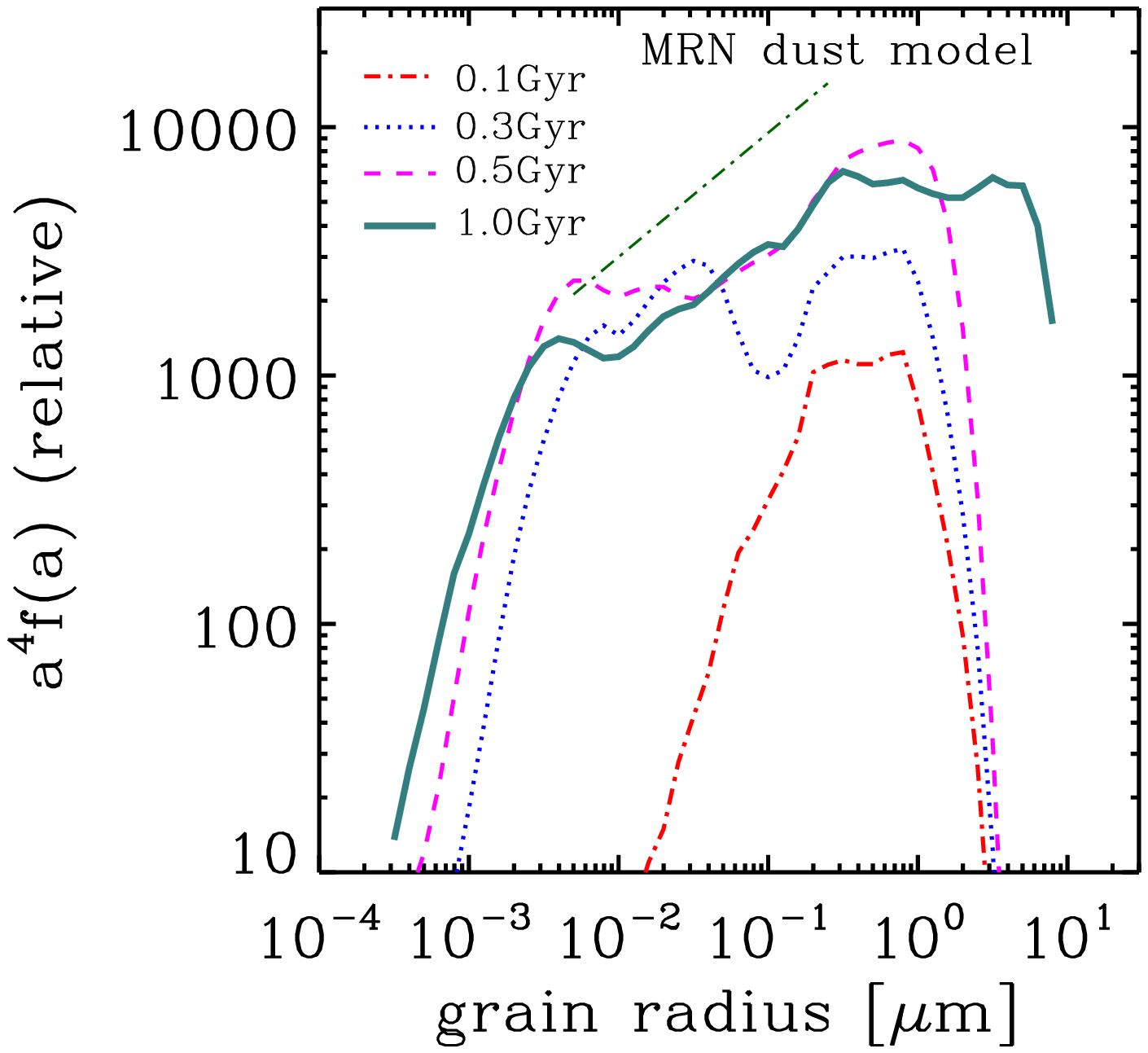}
\caption{
Left: Time evolutions of dust-to-gas mass ratios with grain growth 
(thick solid) and without grain growth (thin solid) 
calculated with $\tau_{\rm SF} = 0.5$ Gyr, 
$\eta_{\rm WNM} = 0.3$, and $\eta_{\rm MC} = 0.7$.
The dashed line draws the time evolution of metallicity 
($M_{Z}/M_{\rm ISM}$), where $M_{Z}$ is the total mass of metals 
in a galaxy.
Right: Grain size distributions for the dust evolution with grain 
growth given by the thick solid line in the left panel.
The dot-dashed, dotted, dashed, and solid lines show the grain
size distributions at $t =$ 0.1, 0.3, 0.5, and 1 Gyr, respectively.
The thin dot-dashed line indicates the MRN size distribution.}
\label{fig2}
\end{figure*}

\section{Dust evolution in high-redshift quasars} \label{sec:qso}

In the last section, we demonstrated that the evolution model of grain 
size distribution reasonably reproduces the MW extinction curve.
Thus, it is a powerful tool with which we can discuss the difference 
in the extinction curve, according to the properties and ages of 
galaxies.
In this section, we apply this model to high-$z$ dusty quasars to 
investigate the evolution of dust grains and extinction curves in 
such young systems.
It should also be emphasized that a mass fraction of each ISM phase 
is important in regulating the size distribution of 
interstellar dust grains, as shown in Section 2.2.

For high-$z$ dusty quasars, it is deduced from their extremely high  
far-infrared luminosities that intense starbursts are ongoing with 
the star formation rates of $\ga$$10^3$ $M_\odot$ yr$^{-1}$
(e.g., Bertoldi et al.\ 2003).
This implies that a huge amount of dense molecular clouds, in which
stars are born, are present in their host galaxies.
In fact, according to the compilation of high-$z$ quasars in 
Calura et al.\ (2014), a mass fraction of H$_2$ molecules to the 
total hydrogen (H + H$_2$) is in the range of 70--97 \%, suggesting 
that a mass fraction of the MC phase is significantly high in these 
quasar hosts.
Therefore, in the calculations of dust evolution in high-$z$ quasars, 
we adopt a two-phase model of WNM and MC with 
$\eta_{\rm WNM} = 0.3$ and $\eta_{\rm MC} = 0.7$.

Figure 2 (left panel) shows the evolutions of dust-to-gas mass ratios 
($M_{\rm d}/M_{\rm ISM}$) and metallicity ($M_Z/M_{\rm ISM}$) calculated by 
our dust evolution model.
Since the star formation rates are found to be very high for high-$z$ 
dusty quasars, we here adopt $\tau_{\rm SF} = 0.5$ Gyr, which is ten 
times shorter than in the case of the MW.
At the initial phases of the galaxy evolution ($t \la 0.2$ Gyr), 
stellar ejecta enrich the ISM with dust grains of $a =$ 0.1--1 $\mu$m 
(see the right panel of Figure 2).
Then, shattering becomes effective at $t \simeq 0.2$ Gyr, producing 
a large number of small grains.
Since the ISM is considerably enriched with metals
($M_Z/M_{\rm ISM} \simeq 0.005 \simeq 0.3$ $Z_\odot$) by this epoch, 
these small grains efficiently grow through the accretion of gaseous 
metals in MCs, enhancing the dust-to-gas mass ratio quickly. 
Around $t = 1$ Gyr, the dust mass produced by grain growth accounts 
for more than 80 \% of the total dust mass, and
the dust-to-gas mass ratio increases up to $\simeq$0.01. 
Given that these high-$z$ quasars have the ISM mass well above 
$M_{\rm ISM} = 10^{10}$ $M_\odot$ (Calura et al.\ 2014), this result
can fully explain the observed masses of dust in excess of 
$10^8$ $M_\odot$.

Figure 2 also presents the evolution of dust-to-gas mass ratio 
without grain growth (thin solid line). 
If grain growth does not work, the dust-to-gas mass ratio reaches 
only $\simeq 1.5 \times 10^{-3}$ at $t=$ 1--2 Gyr.
This value exceeds a lower limit of dust-to-gas mass ratios 
($\simeq$$8 \times 10^{-4}$),
derived by taking the dynamical mass as the total mass of 
the ISM (Kuo \& Hirashita 2012).
However, the dynamical mass includes all components (gas, stars, and
dark matter), so this lower limit is likely to underestimate largely 
the dust-to-gas mass ratio for most of the high-$z$ quasars.
Although we could not completely reject the model without grain 
growth, the model with grain growth more robustly explains the large
dust-to-gas ratio in high-$z$ quasars (see also Valiante et al.\ 2011).

Figure 3 shows the time evolution of the UV extinction curve 
(normalized to the extinction at $\lambda = 0.3$ $\micron$) 
corresponding to the dust evolution with grain growth given in 
Figure 2.
At $t = 0.1$ Gyr, when SNe II predominantly inject large grains into 
the ISM, the extinction curve is flat.
From $t \simeq 0.2$ Gyr, at which small grains are produced due to
shattering, the extinction curve becomes steep, and its shape is 
regulated by the dust processes in the ISM (shattering and grain 
growth) not by the properties of dust supplied by stellar sources.
After that, the UV extinction curve becomes flatter again due to 
the increase in large grains by efficient coagulation in MCs.
At $t=1$ Gyr, the extinction curve has the prominent 2175 \AA~bump, 
but its overall slope is much flatter than the SMC extinction curve, 
which is considered to be valid for low-$z$ quasars.

\begin{figure}
\centering\includegraphics[width=0.42\textwidth]{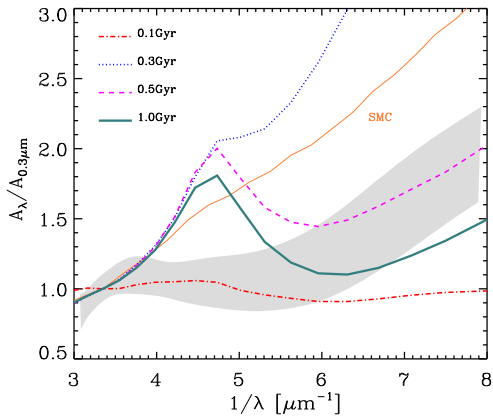}
\caption{Time evolution of the UV extinction curve 
($A_V/A_{0.3 \mu{\rm m}}$) for the dust evolution with grain growth given 
in Figure 2.
Here, the optical constants of graphite are used for carbonaceous 
grains.
The dot-dashed, dotted, dashed, and thick solid lines show the 
extinction curves at $t =$ 0.1, 0.3, 0.5, and 1 Gyr, respectively.
For reference, the SMC extinction curve is depicted by the thin solid 
line, and the range of the extinction curve including the uncertainty 
derived for the quasar SDSS J1048+4637 at $z=6.2$ 
(Maiolino et al.\ 2004) are drawn as the thin hatched region. 
}
\label{fig3}
\end{figure}

\begin{figure}
\centering\includegraphics[width=0.42\textwidth]{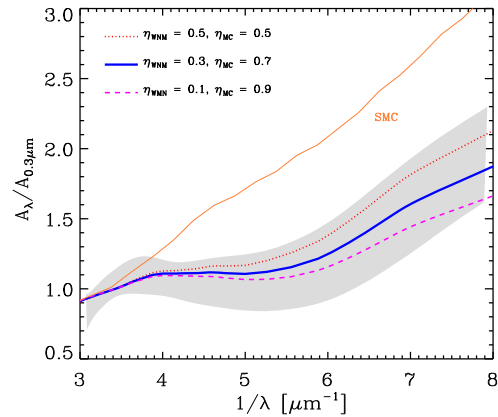}
\caption{
UV extinction curves ($A_V/A_{0.3 \mu{\rm m}}$) at $t = 1$ Gyr derived 
with the optical constants of amorphous carbon for carbonaceous grains.
The dotted, thick solid, and dashed lines show the cases with
$(\eta_{\rm WNM}, \eta_{\rm MC}) = (0.5, 0.5), (0.3, 0.7), (0.1, 0.9)$,
respectively.
The hatched region is the range of the extinction curve including
the uncertainty derived for the quasar SDSS J1048+4637 at $z=6.2$ 
(Maiolino et al.\ 2004). 
The SMC extinction curve is drawn by the thin solid line.}
\label{fig4}
\end{figure}

We note that, if SNe II are only the sources of interstellar dust 
(i.e.\ without grain growth) throughout the galaxy evolution, the 
resulting extinction curves remain too flat, like the one at $t=0.1$ 
Gyr (dot-dashed line) in Figure 3.
Such flat extinction curves would be inconsistent with the $z=6.2$ 
extinction curve (see also Hirashita et al.\ 2008).
Furthermore, it is hard to explain the observed dust content only 
by dust production in SNe II.
Thus, the above results indicate that the high dust-to-gas mass ratio 
and the tendency that extinction curves are moderately flat in 
high-$z$ quasars can be simultaneously explained only by 
introducing a large molecular-cloud fraction which leads to high 
rates of grain growth and coagulation.

It should be noticed here that the calculated extinction curves show 
the prominent 2175-\AA\ bump, which disagrees with the extinction 
curves observed for high-$z$ quasars.
This might be a discrepancy arising from the grain material rather than
from the grain size distribution.
We used graphite to fit the 2175-\AA\ bump, but it has been suggested 
that the carrier of this UV bump is polycyclic aromatic hydrocarbons 
(PAHs) (Joblin, Leger, \& Martin 1992; Weingartner \& Draine 2001).
PAHs are likely to form through photo-fragmentation of small 
amorphous hydrocarbon (Jones 2009) and/or shattering of large 
carbonaceous grains (Seok, Hirashita, \& Asano 2014), while PAHs are 
efficiently destroyed by SN shocks and/or UV radiation from massive 
stars, which is not included in our model. 
The observations of some carbonaceous features including mid-infrared 
PAH emissions in high-$z$ galaxies may give us a clue to 
the properties of carbonaceous dust in the early universe
(e.g.\ Riechers et al.\ 2014).

Because of poor knowledge on the properties of carbonaceous
materials in high-$z$ galaxies, we also calculate the extinction 
curves by adopting the optical constants of amorphous carbon by 
Zubko et al.\ (1996), instead of graphite.
The results are shown in Figure 4, where the extinction curves at 
$t=1$ Gyr are plotted for the cases of 
$(\eta_{\rm WNM}, \eta_{\rm MC}) = (0.5,\, 0.5)$, 
$(0.3,\, 0.7)$, and $(0.1,\, 0.9)$.
For a higher $\eta_{\rm MC}$, the extinction curve is flatter because
more efficient coagulation is realized. 
As is evident from Figure 4, changing the properties of carbonaceous 
materials with a large $\eta_{\rm MC}$ ($\ga 0.5$) leads to a 
successful fit to the high-$z$ quasar extinction curve represented 
by the one in SDSS J1048+4637.

The above results imply that the primary form of carbonaceous grains
in high-$z$ dusty quasars is amorphous carbon, and that the fraction 
of molecular clouds is higher than $\simeq$0.5 in the host galaxies.
Although the origin of the different composition of carbonaceous 
materials in high-$z$ galaxies from those in the local galaxies cannot 
be resolved by our dust evolutin model, the absence of the 
2175-\AA~bump in high-$z$ extinction curves may suggest inefficient 
formation and/or efficient destruction of small size of graphite and 
PAHs.
Furthermore, a high fraction of molecular clouds is the key in 
explaining a high dust-to-gas mass ratio in high-$z$ quasars, 
since molecular clouds are unique sites in which dust mass growth 
by accretion, as well as dust size growth by coagulation, occurs 
efficiently.

\section{Conclusion} \label{sec:conclusion}

We investigated how the massive amounts of dust and the extinction 
curves in high-$z$ dusty quasars can be reproduced in a 
self-consistent manner, and discussed the difference in the ISM 
condition between the Milky Way (MW) and the high-$z$ quasars. 
First, we found that the MW extinction curve is successfully 
reproduced by our dust evolution model with a three-phase ISM 
including the phase of molecular clouds.
We also showed that our model can simultaneously account for the 
large amount of dust and the peculiar extinction curves in high-$z$ 
quasars if a large fraction of the ISM mass is in molecular clouds.
These results demonstrate that grain growth and coagulation in 
molecular clouds are the key processes to detemine the properties 
of interstellar dust both in the MW and high-$z$ galaxies.
In addition, the optical properties of carbonaceous dust are needed to 
be changed:
graphite is used for the MW, while amorphous carbon is applied for 
the high-$z$ quasars.
This indicates the difference in the property of carbonaceous dust 
between the MW and the high-$z$ galaxies.

\section*{Acknowledgments}

We are grateful to the anonymous referee for critical comments. 
We are also grateful to Y. Matsuoka and T. Minezaki for useful 
comments.
We thank R. Maiolino for kindly providing us with the data on 
the extinction curve of SDSS J1048+4637.
TN and TTT are supported by JSPS KAKENHI 
(22684004, 23224004, 23340046, 26400223).
HH thanks the support from the Ministry of Science and Technology
(MoST) grant 102-2119-M-001-006-MY3.



\bsp

\label{lastpage}


\begin{thebibliography}{99}

\bibitem[\protect\citeauthoryear{Asano et al.}{2013a}]{asano13a}
Asano R. S., Takeuchi T. T., Hirashita H., Inoue A. K.,
2013a, Earth Planets and Space, 65, 213

\bibitem[\protect\citeauthoryear{Asano et al.}{2013b}]{asano13b}
Asano R. S., Takeuchi T. T., Hirashita H., Nozawa T.,
2013b, MNRAS, 432, 637

\bibitem[\protect\citeauthoryear{Asano et al.}{2014}]{asano14}
Asano R. S., Takeuchi T. T., Hirashita H., Nozawa T., 
2014, MNRAS, 440, 134

\bibitem[\protect\citeauthoryear{Bertoldi et al.}{2003}]{bertoldi}
Bertoldi F., Carilli C. L., Cox P., Fan X., Strauss M. A., Beelen A.,
Omount A., Zylka R., 2003, A\&A, 406, L55

\bibitem[\protect\citeauthoryear{Bianchi \& Schneider}{2007}]{bianchi}
Bianchi S., Schneider R., 2007, MNRAS, 378, 973

\bibitem[\protect\citeauthoryear{Calura et al.}{2014}]{calura14}
Calura F., Gilli R., Vignali C., Pozzi F., Pipino A., Matteucci F., 
2014, MNRAS, 438, 2765

\bibitem[\protect\citeauthoryear{Chokshi, Tielens, \& Hollenbach}{1993}]{chokshi}
Chokshi A., Tielens A. G. G. M., Hollenbach, D., 1993, ApJ, 407, 806

\bibitem[\protect\citeauthoryear{Draine \& Lee}{1984}]{dralee}
Draine B. T., Lee H. M., 1984, ApJ, 285, 89

\bibitem[\protect\citeauthoryear{Dwek \& Scalo}{1980}]{dwek80}
Dwek E., Scalo, J. M., 1980, ApJ, 239, 193

\bibitem[\protect\citeauthoryear{Dwek et al}{2007}]{dwek07}
Dwek E., Galliano F., Jones A. P., 2007, ApJ, 662, 927

\bibitem[\protect\citeauthoryear{Dwek \& Cherchneff}{2011}]{dwek11}
Dwek E., Cherchneff I., 2011, ApJ, 727, 63

\bibitem[\protect\citeauthoryear{Fitzpatrick \& Massa}{2007}]{fitzpatrick}
Fitzpatrick E. L., Massa D., 2007, ApJ, 663, 320

\bibitem[\protect\citeauthoryear{Gall et al.}{2011a}]{gall11a}
Gall C., Andersen A. C., Hjorth J., 2011a, A\&A, 528, 13

\bibitem[\protect\citeauthoryear{Gall et al.}{2011b}]{gall11b}
Gall C., Andersen A. C., Hjorth J., 2011b, A\&A, 528, 14

\bibitem[\protect\citeauthoryear{Gallerani et al.}{2010}]{gallerani}
Gallerani S., et al., 2010, A\&A, 523, 85

\bibitem[\protect\citeauthoryear{Gaskwell et al.}{2004}]{gaskwell}
Gaskwell C. M., Goosmann R. W., Antonucci R. R. J., Whysong D. H.,
2004, ApJ, 616, 147

\bibitem[\protect\citeauthoryear{Hirashita et al.}{2005}]{hirashita05}
Hirashita H., Nozawa T., Kozasa T., Ishii T. T., Takeuchi T.~T.,
2005, MNRAS, 357, 1077

\bibitem[\protect\citeauthoryear{Hirashita et al.}{2008}]{hirashita08}
Hirashita H., Nozawa T., Takeuchi T. T., Kozasa T.,
2008, MNRAS, 384, 1725

\bibitem[\protect\citeauthoryear{Hirashita \& Yan}{2009}]{hirashita09b}
Hirashita H., Yan H., 2009, MNRAS, 394, 1061

\bibitem[\protect\citeauthoryear{Hirashita}{2011}]{hirashita11}
Hirashita H., Kuo T.-M., 2011, MNRAS, 416, 1340

\bibitem[\protect\citeauthoryear{Hirashita \& Voshchinnikov}{2014}]{hirashita14}
Hirashita H., Voshchinnikov N. V., 2014, MNRAS, 437, 1636

\bibitem[\protect\citeauthoryear{Hjorth et al.}{2013}]{hjorth}
Hjorth J., Vreeswijk P. M., Gall C., Watson D., 2013, ApJ, 768, 173

\bibitem[\protect\citeauthoryear{Hopkins et al.}{2004}]{hopkins04}
Hopkins P. F., et al., 2004, AJ, 128, 1112

\bibitem[\protect\citeauthoryear{Inoue}{2011}]{inoue}
Inoue A. K., 2011, Earth Planets and Space, 63, 1

\bibitem[\protect\citeauthoryear{Joblin, Leger, \& Martin}{1992}]{joblin}
Joblin C., Leger A., Martin P., 1992, ApJ, 393, L79

\bibitem[\protect\citeauthoryear{Jones, Tielens \& Hollenbach}{1996}]{jones96}
Jones A. P., Tielens A. G. G. M., Hollenbach D. J.,
1996, ApJ, 469, 740

\bibitem[\protect\citeauthoryear{Jones}{2009}]{jones09}
Jones A. P., 2009, in Henning Th., Gr\"{u}n E., Steinacker J., 
eds, ASP Conf.\ Ser.\ Vol.\ 414, Cosmic Dust -- 
Near and Far. Astron.\ Soc.\ Pac., San Francisco, p.\ 473

\bibitem[\protect\citeauthoryear{Kuo \& Hirashita}{2012}]{kuo}
Kuo T.-M., Hirashita H., 2012, MNRAS, 424, L34

\bibitem[\protect\citeauthoryear{Maiolino et al.}{2004}]{maiolino}
Maiolino R., Schneider R., Oliva E., Bianchi S., Ferrara A.,
Mannucci F., Pedani M., Roca Sogorb, M.,
2004, Nature, 431, 533

\bibitem[\protect\citeauthoryear{Mathis, Rumpl \& Nordsieck}{1977}]{MRN}
Mathis J. S., Rumpl W., Nordsieck, K. H., 1977, ApJ, 217, 425

\bibitem[\protect\citeauthoryear{Mattsson}{2011}]{mattsson}
Mattsson L., 2011, MNRAS, 414, 781

\bibitem[\protect\citeauthoryear{Micha{\l}owski et al.}{2010}]{michalowski}
Michalowski M. J., Murphy E. J., Hjorth J., Watson D., Gall C.,
Dunlop J. S., 2010, A\&A, 522, 15

\bibitem[\protect\citeauthoryear{Nozawa et al.}{2003}]{nozawa03}
Nozawa T., Kozasa T., Umeda H., Maeda K., Nomoto K., 
2003, ApJ, 598, 785

\bibitem[\protect\citeauthoryear{Nozawa et al.}{2007}]{nozawa07}
Nozawa T., Kozasa T., Habe A., Dwek E., Umeda H., Tominaga N.,
Maeda K., Nomoto, K., 2007, ApJ, 666, 955 

\bibitem[\protect\citeauthoryear{Nozawa \& Fukugita}{2013}]{nozawa13}
Nozawa T., Fukugita, M., 2013, ApJ, 770, 27

\bibitem[\protect\citeauthoryear{Pipino et al.}{2011}]{pipino11}
Pipino A., Fan X. L., Matteucci F., Calura F., Silva L., Granato G.,
Maiolino, R., 2011, A\&A, 525, 61

\bibitem[\protect\citeauthoryear{Richards et al.}{2003}]{richards03}
Richards G. T., et al., 2003, AJ, 126, 1131

\bibitem[\protect\citeauthoryear{Riechers et al.}{2014}]{reichers14}
Riechers D. A., et al., 2014, ApJ, 786, 31

\bibitem[\protect\citeauthoryear{Seok, Hirashita, \& Asano}{2014}]{seok14}
Seok J. Y., Hirashita H., Asano, R. S., 2014, MNRAS, 439, 2186

\bibitem[\protect\citeauthoryear{Valiante et al.}{2009}]{valiante09}
Valiante R., Schneider R., Bianchi S., Andersen A. C.,
2009, MNRAS, 397, 1661

\bibitem[\protect\citeauthoryear{Valiante et al.}{2011}]{valiante11}
Valiante R., Schneider R., Salvadori S., Bianchi S.,
2011, MNRAS, 416, 1916

\bibitem[\protect\citeauthoryear{Weingartner \& Draine}{2001}]{WD01}
Weingartner J. C., Draine B. T., 2001, ApJ, 548, 296

\bibitem[\protect\citeauthoryear{Whittet}{2003}]{whittet03}
Whittet D. C. B., 2003, Dust in the Galactic Environments, 2nd edn.
IoP Publishing, Bristol, P. 76

\bibitem[\protect\citeauthoryear{Willott et al.}{2007}]{willott07}
Willott C. J., et al., 2007, AJ, 134, 2435

\bibitem[\protect\citeauthoryear{Yamasawa et al.}{2011}]{yamasawa}
Yamasawa D., Habe A., Kozasa T., Nozawa T., Hirashita H., Umeda H., 
Nomoto K., 2011, ApJ, 735, 44

\bibitem[\protect\citeauthoryear{Yan et al.}{2004}]{yan}
Yan H., Lazarian A., Draine B. T., 2004, ApJ, 616, 895

\bibitem[\protect\citeauthoryear{Zhukovska, Gail, \& Trieloff}{2008}]{zhukovska}
Zhukovska S., Gail H. P., Trieloff M., 2008, A\&A, 479, 453

\bibitem[\protect\citeauthoryear{Zubko et al.}{1996}]{zubko96}
Zubko V. G., Mennella V., Colangeli L., Bussoletti E., 
1996, MNRAS, 282, 1321

\bibitem[\protect\citeauthoryear{Zubko, Dwek, \& Arendt}{2004}]{zubko04}
Zubko V., Dwek E., Arendt R. G., 2004, A\&A, ApJS, 152, 211


\end{thebibliography}
\end{document}